\documentclass[10pt]{article}
\usepackage{latexsym}
\usepackage{amssymb}
\usepackage{amsmath}
\usepackage{amscd}
\usepackage{amsthm}
\usepackage[left=2cm,top=2.5cm,right=2.5cm,bottom=1.5cm]{geometry}
\usepackage[dvips]{graphicx}
\usepackage{subcaption}
\usepackage{caption}
\usepackage{epstopdf}
\usepackage{hyperref}
\begin{document}
\begin{center}
\large{\bf{Constraints on The Dark Energy Equation of State And The Deceleration Parameter From Recent Cosmic Observations}} \\
\vspace{10mm}
\normalsize{Hassan Amirhashchi}\\
\vspace{5mm} \normalsize{Department of Physcis, Islamic Azad University, Mahshahr Branch, Mahshahr, Iran \\
E-mail:h.amirhashchi@mahshahriau.ac.ir; hashchi@yahoo.com}  \\

\end{center}
\vspace{10mm}
\begin{abstract}
We study the constraints on dark energy equation of state $\omega^{X}$ and the deceleration parameter $q$ from the recent observational data including Hubble data and the cosmic microwave background (CMB) radiation by using a model-independent deceleration parameter $q(z)=1/2-a/(1+z)^b$ and dark energy equation of state $\omega^{X}=\omega_{0}+\omega_{1}z/(1+z)$ in the scope of anisotropic bianchi type I space-time.
For the cases of Hubble dataset, CMB data, and their combination, our results indicate that the constraints on transition redshift $z_{\ast}$
are $0.62^{+1.45}_{-0.56}$, $0.34^{+0.13}_{-0.06}$, and $0.60^{+0.20}_{-0.10}$ respectively.
\end{abstract}
 \smallskip
Keywords : Bianchi Type I Model, Dark Energy, Experimental Tests \\
PACS number: 98.80.Es, 98.80-k, 95.36.+x
\section{Introduction}
According to the recent cosmic observations our universe is experiencing an accelerating phase of expansion at the present time \cite{ref1}-\cite{ref5}. Since the ordinary barionic matter (energy) can only produces an attractive force, there should necessarily  an exotic form of energy with negative pressure called ``dark energy" (DE) be exist in order to drive the observed cosmic acceleration \cite{ref5}-\cite{ref8}. Moreover, based on the recent observations we live in a nearly spatially flat Universe composed of approximately $4\%$ baryonic matter, $22\%$ dark matter and $74\%$ dark energy. A natural candidate for dark energy is cosmological constant $\Lambda$, which has the equation of state $\omega_{\Lambda}=p_{\Lambda}/\rho_{\Lambda} = -1$. However, although cosmological constant can explain the present cosmic acceleration, it would encounter some serious theoretical problems, such as the fine-tuning and the coincidence problems. Another possible candidates for dark energy are
the dynamic dark energy models provided by scalar fields such as quintessence $-\frac{1}{3}>\omega>-1$ \cite{ref9}-\cite{ref14}, phantom ($\omega^{X}<-1$) \cite{ref15}, quintom ($\omega^{X}<-\frac{1}{3}$) \cite{ref16}, k-essence \cite{ref17, ref18}, Chaplygin gas as well as generalized Chaplygin gas models \cite{ref19, ref20}, and etc.\\

We also can study dark energy through an almost model-independent way. In this approach we parameterize the dark energy EoS parameter by giving the concrete form of the equation of state of dark energy directly, such as $\omega(z)=\omega_{0}+\omega_{1}z$ \cite{ref21}, $\omega(z)=\omega_{0}+\omega_{1}\frac{z}{1+z}$ \cite{ref22, ref23}, etc. Studies of dark energy as a function of redshift, by using this method, show that though the cosmological constant is not ruled out in 1$\sigma$ region, the current constraints favors a dynamical dark energy.
However, as noted in ref \cite{ref24} the rapid changed of EoS parameter i.e $|\frac{\partial \omega}{\partial z}|\ll 1$ is ruled out.
Moreover, it is a well established fact that the universe is accelerated expansion at present (dark energy dominated) and decelerated expansion in the past (dark matter dominated). Therefore, the deceleration parameter (DP) should not be a constant but time (or redshift) variable which is positive in the past and changes into negative at $z_{\ast}\sim 0.5$ \cite{ref24}-\cite{ref26}.
Considering the idea of variable DP, the parameterized decelerated parameter is present in almost model independent way by giving a concrete form of
decelerated parameters.\\

The high symmetry involved in FLRW models requires a very high degree of fine tuning of initial conditions which is extraordinary improbable and hence FLRW models are infinitely improbable in the space of all possible cosmologies \cite{ref27}. Although there are an increasing interest in the study of DE in the scop of anisotropic space-times, there is not a reference paper including the present observational constraints on the observable parameters such as $q$, $\omega$, $z_{\ast}$, etc. Therefore, still authors compare their results by those obtained on the bases of the standard cosmological models. Motivated the situation discussed above, in this letter instead of FLRW metric, an anisotropic space-time namely Bianchi type I metric is used in order to obtain more general results. Up to our knowledge, such kind of calculations has not been done yet. First we derive the general form of the EoS parameter in Bianchi type I space-time and then we use the Hubble dataset, CMB data, and their combination to put the observational constraints on the model parameters. I is worth to mention that because of technical difficulties in minimizing $\chi^{2}$ the results obtained in this work are not so general and more work is needed to extend these results for example by minimizing $\chi^{2}_{SneIa}$.
\section{Dark Energy Equation of State}
We consider the Bianchi type I space-time in the orthogonal form
as
\begin{equation}
\label{eq1}
ds^{2} = -dt^{2} + A^{2}(t)dx^{2}+B^{2}(t)dy^{2}+C^{2}(t)dz^{2},
\end{equation}
where $A(t), B(t)$ and $C(t)$ are functions of time only. \\

The Einstein's field equations ( in gravitational units $8\pi G = c = 1 $) read as
\begin{equation}
\label{eq2} R^{i}_{j} - \frac{1}{2} R g^{i}_{j} = T^{(m) i}_{j} +
T^{(X) i}_{j},
\end{equation}
where $T^{(m) i}_{j}$ and $T^{(X) i}_{j}$ are the energy momentum tensors of barotropic matter and dark energy,
respectively. These are given by
\[
  T^{m i}_{j} = \mbox{diag}[-\rho^{m}, p^{m}, p^{m}, p^{m}],
\]
\begin{equation}
\label{eq3} ~ ~ ~ ~ ~ ~ ~ ~  = \mbox{diag}[-1, \omega^{m}, \omega^{m}, \omega^{m}]\rho^{m},
\end{equation}
and
\[
 T^{X i}_{j} = \mbox{diag}[-\rho^{X}, p^{X}, p^{X}, p^{X}],
\]
\begin{equation}
\label{eq4} ~ ~ ~ ~ ~ ~ ~ ~ ~ ~ ~ ~ ~ ~ = \mbox{diag}[-1, \omega^{X}, \omega^{X},
\omega^{X}]\rho^{X},
\end{equation}
where $\rho^{(m)}$ and $p^{(m)}$ are, respectively the energy density and pressure of the perfect fluid
component or ordinary baryonic matter while $\omega^{m} = p^{m}/\rho^{m}$ is its EoS parameter. Similarly,
$\rho^{X}$ and $p^{X}$ are, respectively the energy density and pressure of the DE component while
$\omega^{X} = p^{X}/\rho^{X}$ is the corresponding EoS parameter. We assume the four velocity vector
$u^{i} = (1, 0, 0, 0)$ satisfying $u^{i}u_{j} = -1$. \\

In a co-moving coordinate system ($u^{i} = \delta^{i}_{0}$), Einstein's field equations (\ref{eq2}) with
(\ref{eq3}) and (\ref{eq4}) for B-I metric (\ref{eq1}) subsequently lead to the following system of equations:
\begin{equation}
\label{eq5} \frac{\ddot{B}}{B}+\frac{\ddot{C}}{C}+\frac{\dot{B}\dot{C}}{BC}=-\omega^{m}\rho^{m}-\omega^{X}\rho^{X},
\end{equation}
\begin{equation}
\label{eq6} \frac{\ddot{A}}{A}+\frac{\ddot{C}}{C}+\frac{\dot{A}\dot{C}}{AC}=-\omega^{m}\rho^{m}-\omega^{X}\rho^{X},
\end{equation}
\begin{equation}
\label{eq7}\frac{\ddot{A}}{A}+\frac{\ddot{B}}{B}+\frac{\dot{A}\dot{B}}{AB}=-\omega^{m}\rho^{m}-\omega^{X}\rho^{X},
\end{equation}
\begin{equation}
\label{eq8} \frac{\dot{A}\dot{B}}{AB}+\frac{\dot{A}\dot{C}}{AC}+\frac{\dot{B}\dot{C}}{BC}=\rho^{m}+\rho^{X}.
\end{equation}
If we consider $a=(ABC)^{\frac{1}{3}}$ as the average scale factor of Bianchi type I model, then
the generalized mean Hubble's parameter $H$ defines as
\begin{equation}
\label{eq9} H = \frac{\dot{a}}{a} = \frac{1}{3}\left(\frac{\dot{A}}{A} + \frac{\dot{B}}{B} + \frac{\dot{C}}{C}\right).
\end{equation}

The Bianchi identity $G^{;j}_{ij} = 0$ leads to  $T^{;j}_{ij} = 0$. Therefore, the continuity equation
for dark energy and baryonic matter can be written as
\begin{equation}
\label{eq10} \dot{\rho}^{m} + 3H(1 + \omega^{m})\rho^{m} + \dot{\rho}^{X} + 3H(1 + \omega^{X})\rho^{X} = 0.
\end{equation}
Solving eqs. (\ref{eq5}-(\ref{eq8}) one can find \cite{ref28}
\begin{equation}
\label{eq11} A(t)=a_{1}a~ exp(b_{1}\int a^{-3}dt),
\end{equation}
\begin{equation}
\label{eq12} B(t)=a_{2}a~ exp(b_{2}\int a^{-3}dt),
\end{equation}
and
\begin{equation}
\label{eq13} C(t)=a_{3}a~ exp(b_{3}\int a^{-3}dt),
\end{equation}
where
\[
a_{1}a_{2}a_{3}=1,~~~~~~~b_{1}+b_{2}+b_{3}=0.
\]

Using eqs. (\ref{eq11})-(\ref{eq13}) in eqs.
(\ref{eq5})-(\ref{eq8}) we can write the analogue of the Friedmann
equation as
\begin{equation}
\label{eq14} \left(\frac{\dot{a}}{a}\right)^{2}=\frac{\rho}{3}+ K
a^{-6},
\end{equation}
and
\begin{equation}
\label{eq15}
2\left(\frac{\ddot{a}}{a}\right)=-\frac{1}{3}(\rho+3p).
\end{equation}
Here $\rho = \rho^{m} + \rho^{de}$, $p = p^{m} + p^{de}$ and
$K=b_{1}b_{2}+b_{1}b_{3}+b_{2}b_{3}$. Note that $K$ denotes the
deviation from isotropy e.g. $K=0$ represents flat FLRW universe.
Thus, when the universe is sufficiently large, almost at the present time, the space-time
(\ref{eq1}) behaves like a flat FLRW universe.\\

Moreover, if there is no interaction between dark energy and cold dark matter(CDM) with
$\omega_{m}=0$, then one can write the conservation equation
(\ref{eq10}) for CDM and dark energy separately as
separately as
\begin{equation}
\label{eq16} \dot{\rho}^{X} + 3H(1 + \omega^{X})\rho^{X} = 0,
\end{equation}
and
\begin{equation}
\label{eq17} \dot{\rho}^{m} + 3H\rho^{m}=0.
\end{equation}
Eq.(\ref{eq16}) leads to
\begin{equation}
\label{eq18} \rho^{m}=\rho_{0}^{m}a^{-3}.
\end{equation}
Using eqs. (\ref{eq14}), (\ref{eq18}) in eqs. (\ref{eq7}) and
(\ref{eq8}) we obtain the energy density and pressure of dark
fluid as
\begin{equation}
\label{eq19} \rho^{X}=3\left(\frac{H}{H_{0}}\right)^{2}-3Ka^{-6}-\rho_{0}^{m}a^{-3}
\end{equation}
and
\begin{equation}
\label{eq20} p^{X}=-2\frac{\ddot{a}}{a}-\left(\frac{H}{H_{0}}\right)^{2}-La^{-6},
\end{equation}
respectively. Here, $L=b_{2}^{2}+b_{3}^{2}+b_{2}b_{3}$ is a positive constant (Note
that $K+L=0$). Therefore, the equation of state parameter (EoS) of
DE in it's general form is given by
\begin{equation}
\label{eq21}
\omega^{X}=\frac{p^{X}}{\rho^{X}}=\frac{2q-1-La^{-6}\left(\frac{H}{H_{0}}\right)^{-2}}
{3+3La^{-6}\left(\frac{H}{H_{0}}\right)^{-2}-3\Omega_{0}^{m}a^{-3}},
\end{equation}
where $q=-\frac{\ddot{a}}{aH^{2}}$ is the deceleration parameter and
$\Omega_{0}^{m}$ is the current value of matter density.\\
\section{Experimental Tests}
In this section, by using $\chi^{2}$ method we investigate the constraints on the
parameters $\omega^{X}$, $ q$, and $z_{\ast}$. To do so we utilize the recent
observational data including Hubble parameter and the cosmic microwave background (CMB)
radiation. Note that as mentioned by Kumatsu et al \cite{ref29} the baryonic oscillation (BAO) distance ratio is
not applicable in non-FLRW based models.\\

First of all, from eq. (\ref{eq21}) we obtain the expression for the Hubble rate, of the form
\begin{equation}
\label{eq22}
E^{2}(z)=\frac{H^{2}}{H_{0}^{2}}=\frac{L(1+3\omega^{X})(1+z)^{-6}}{(2q-1)+3\omega^{X}(\Omega^{m}_{0}(1+z)^{3}-1)}.
\end{equation}
In our study, following Xu L X et al. \cite{ref30}, the deceleration parameter is considered to be
\begin{equation}
\label{eq23}
q(z)=\frac{1}{2}-\frac{a}{(1+z)^{b}},
\end{equation}
where $a$ and $b$ are constants determined by the recent observational constraints.
From this equation it is obvious that for $z\gg 1$, $q\to \frac{1}{2}$ which is corresponding
to matter dominated era whereas for $z=0$, the current value of deceleration parameter is obtained as
$q_{0}=\frac{1}{2}-a$. In view of eq. (\ref{eq23}), the Hubble parameter is written in the form
\begin{equation}
\label{eq24}
H(z)=H_{0}(1+z)^{\frac{3}{2}}\mbox{exp}\left[a((1+z)^{-b}-1)/b\right],
\end{equation}
We also consider the following simple expression for the equation of state
\begin{equation}
\label{eq25}
    \omega^{X}=\omega_{0}+\omega_{1}\frac{z}{1+z},
\end{equation}
where $\omega_{0}$ and $\omega_{1}$ are constants. Using eqs. (\ref{eq23}) and (\ref{eq25})
one can easily re-write eq. (\ref{eq22}) as a function of $z$ only.\\

As far as we are dealing with the experimental $H(z)$ test, it has been suggested by Jimenez et. al. \cite{ref31} to use
the quantity $\frac{dz}{dt}$ which is called {\it differential age}. In this method, one can directly analyze the $H(z)$ data without
passing through the luminosity distance as we usually do in the case of Supernovae Ia.
Using the differential ages of passively-evolving galaxies from the Gemini Deep
Deep Survey (GDDS)\cite{ref32} and archival data \cite{ref33, ref34}, first, a set of nine values of $H(z)$ in the range of $0 \leq z \leq 1.8$ has been reported \cite{ref35}. Table 1 shows the most recent data set includes 14 values.\\

We minimize the following reduced $\chi^{2}$ \cite{ref36} to constrain the parameters
$\omega^{X}$ and $ q$ for fixed $H_{0}$ at $71$ and $\Omega^{m}_{0}=0.27$.
\begin{equation}
\label{eq26}
\chi^{2}_{Hub}(\Omega^{X}, q; H_{0})=\frac{1}{\nu}\sum_{i=1}^{14}\frac{[H^{th}(z_{i}|\Omega^{X}, q; H_{0})-H^{obs}(z_{i})]^{2}}{\sigma^{2}_{Hub}(z_{i})},
\end{equation}
where $H^{obs}$ are the values of Table 1 and $\nu = 13$.\\
\begin{table}[ht]
\caption{The cosmological data at 1$\sigma$ error for $H(z)$ expressed in $s^{-1}MPc^{-1}Km$.}
\centering
\begin{tabular}{|ccccccccccccccc|}
\hline{\smallskip}
$z$ & 0.00 & 0.10 & 0.17 & 0.24 & 0.27 & 0.40 & 0.43 & 0.48 & 0.88 & 0.90 & 1.30 & 1.43 & 1.53 & 1.75 \\[0.5ex]
\hline
$H(z)$ & 72 & 69 & 83 & 79.69 & 77 & 95 & 86.45 & 97 & 90 & 117 & 168 & 177 & 140 & 220\\
$1\sigma~error$ & $\pm$8 & $\pm$12& $\pm$8 & $\pm$4.61 & $\pm$14 & $\pm$17& $\pm$5.96 & $\pm$60& $\pm$40& $\pm$23& $\pm$17 & $\pm$18 & $\pm$14 & $\pm$40 \\[1ex]
\hline
\end{tabular}
\label{table:nonlin}
\end{table}

To constraint our models we have also used the CMB. In this case, by using the shift parameter $R$
one can constraints the model parameters by minimizing
\begin{equation}
\label{eq27}
\chi_{CMB}^{2}=\frac{[R^{th}-R^{obs}]^{2}}{(0.019)^{2}}.
\end{equation}
Here $R^{obs}=1.725\pm0.018$ \cite{ref29} and $R^{th}$ is given by
\begin{equation}
\label{eq28}
R^{th}=(\Omega_{0}^{m})^\frac{1}{2}\int_{0}^{z_{CMB}}\frac{d\acute{z}}{E(\acute{z})},
\end{equation}
where $z_{CMB}=1091.3$.\\
The constraints on model parameters can be obtained by minimizing $\chi^{2}_{Hub}+\chi^{2}_{CMB}$.
First we use eq. (\ref{eq24}) in eqs. (\ref{eq26}) and (\ref{eq27}) to find constraints on $a$, $b$, and $q$, then we apply
eqs. (\ref{eq26}) and (\ref{eq27}) on eq. (\ref{eq22}) using the results from previous calculations.
by ftting the observational Hubble data, CMB, and their combination respectively, we obtained the
values of minimum $\chi^{2}$, best fit parameters $a$, $b$ $\omega_{0}$ and $\omega_{1}$;
transition redshifts $z_{\ast}$ , and current deceleration parameter $q_{0}$.
Our results are listed in Table 2.\\

\begin{table}[ht]
\caption{The best fit parameters with 1$\sigma$ error.}
\centering
\begin{tabular}{|c|c|c|c|c|c|c|c|}
\hline{\smallskip}
observational data & $\chi^{2}_{min}$ & $a$ & $b$ & $\omega_{0}$ & $\omega_{1}$ & $q_{0}$ & $z_{\ast}$\\[0.5ex]
\hline
Hubble data & 9.02 & $0.97^{+1.26}_{-0.56}$ & $1.37^{+0.67}_{-0.25}$ & $-1.08^{+0.78}_{-1.02}$ & $0.83^{+0.27}_{-1.13}$ & $-0.47^{+0.68}_{-0.68}$ & $0.62^{+1.45}_{-0.56}$ \\
CMB data & 161.21 & $1.43^{+0.67}_{-1.43}$ & $2.10^{+0.89}_{-1.60}$ & $-1.42^{+0.92}_{-0.29}$ & $1.25^{+0.25}_{-1.25}$ & $-0.93^{+1.46}_{-48}$ & $0.34^{+0.13}_{-0.06}$  \\

Combination & 168.34 & $1.26^{+0.15}_{-1.25}$ & $1.96^{+0.50}_{-1.89}$ & $-0.98^{+0.88}_{-0.22}$ & $0.43^{+0.17}_{-3.13}$ & $-0.76^{+0.46}_{-0.26}$ & $0.60^{+0.20}_{-0.10}$ \\[1ex]
\hline
\end{tabular}
\label{table:nonlin}
\end{table}
The $1\sigma$ contours of parameters $a$ and $b$, the variation of the decelaration parameter $q$, and the
$1\sigma$ contours of parameters $\omega_{1}$ and $\omega_{0}$ are depicted in figures $1, 2$, and $3$ respectively.
All these Figures show that the combined constraint is stricter than independent constraints given
by CMB dataset or the Hubble data.\\
\begin{figure}
        \centering
        \begin{subfigure}[b]{0.3\textwidth}
                \centering
                \includegraphics[width=\textwidth]{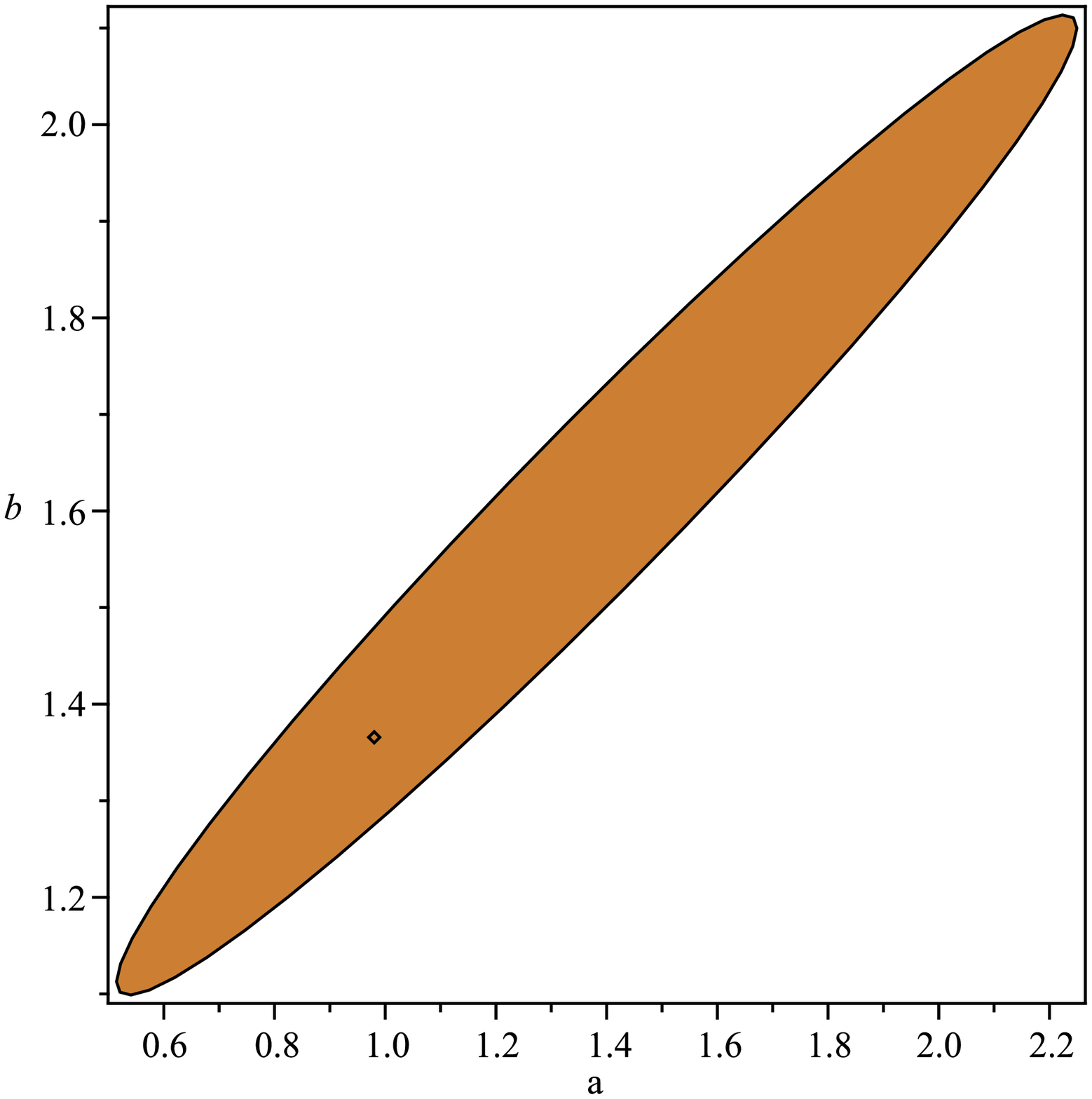}
                \caption{Hubble}
                \label{fig:Hubble}
        \end{subfigure}%
        ~ 
        \begin{subfigure}[b]{0.3\textwidth}
                \centering
                \includegraphics[width=\textwidth]{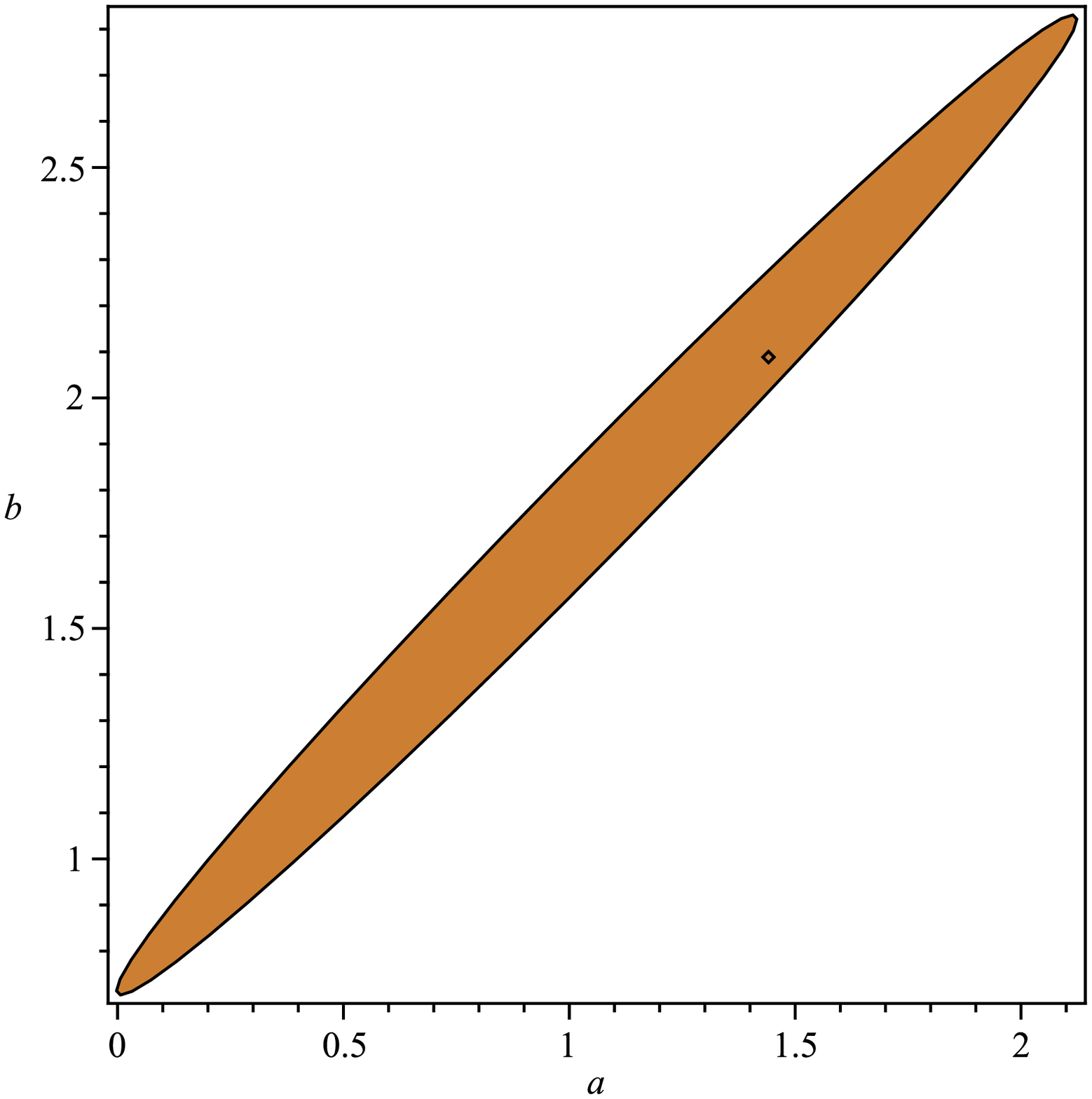}
                \caption{CMB}
                \label{fig:CMB}
        \end{subfigure}
        ~ 
        \begin{subfigure}[b]{0.3\textwidth}
                \centering
                \includegraphics[width=\textwidth]{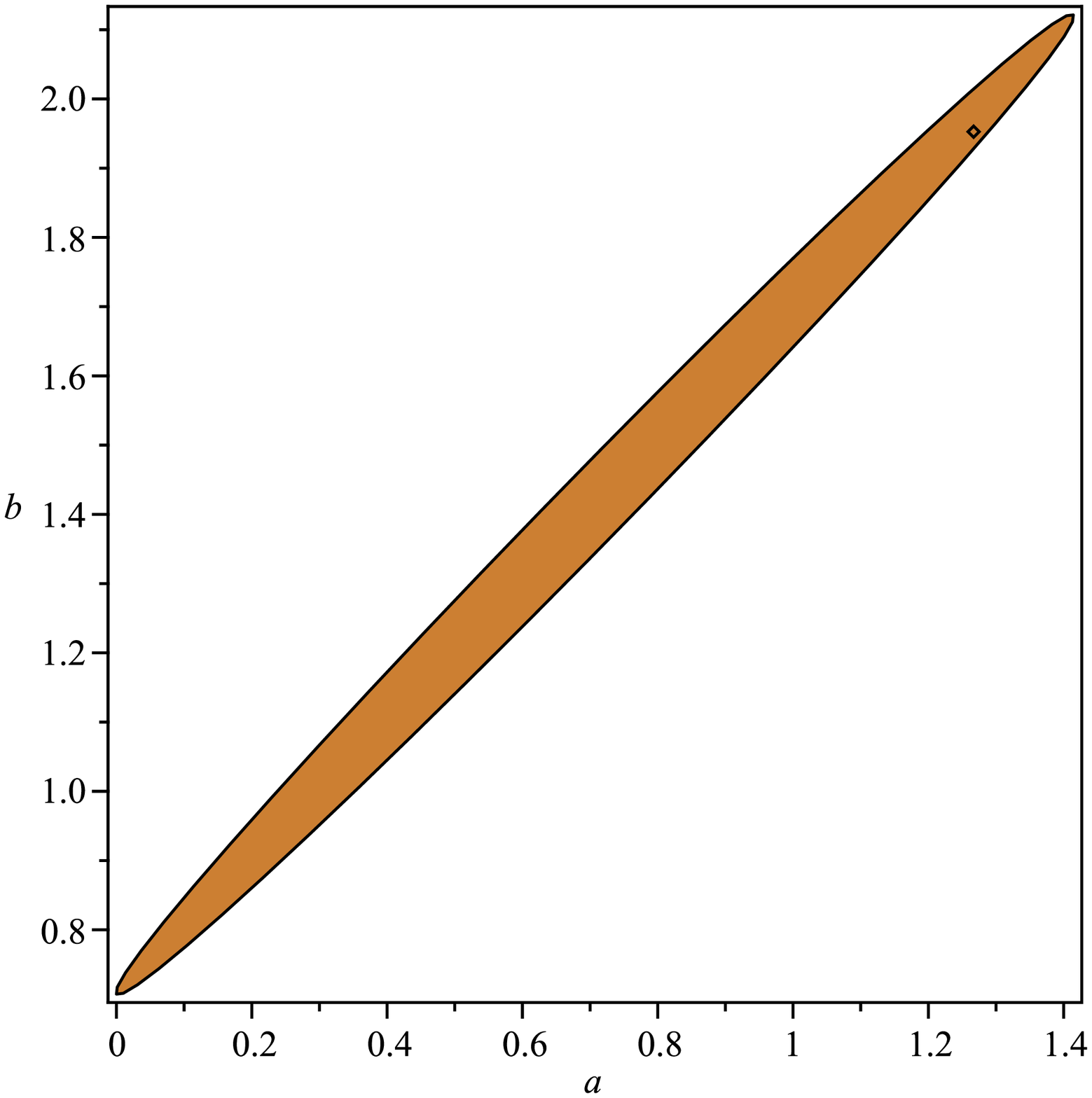}
                \caption{H+CMB}
                \label{fig:H+CMB}
        \end{subfigure}
        \caption{Plots of the parameters $a$ and $b$ with $1\sigma$ confidence level according to: (a) Hubble, (b) CMB, and (c) H + CMB combination data}\label{fig:Observational data}
\end{figure}
\begin{figure}
        \centering
        \begin{subfigure}[b]{0.3\textwidth}
                \centering
                \includegraphics[width=\textwidth]{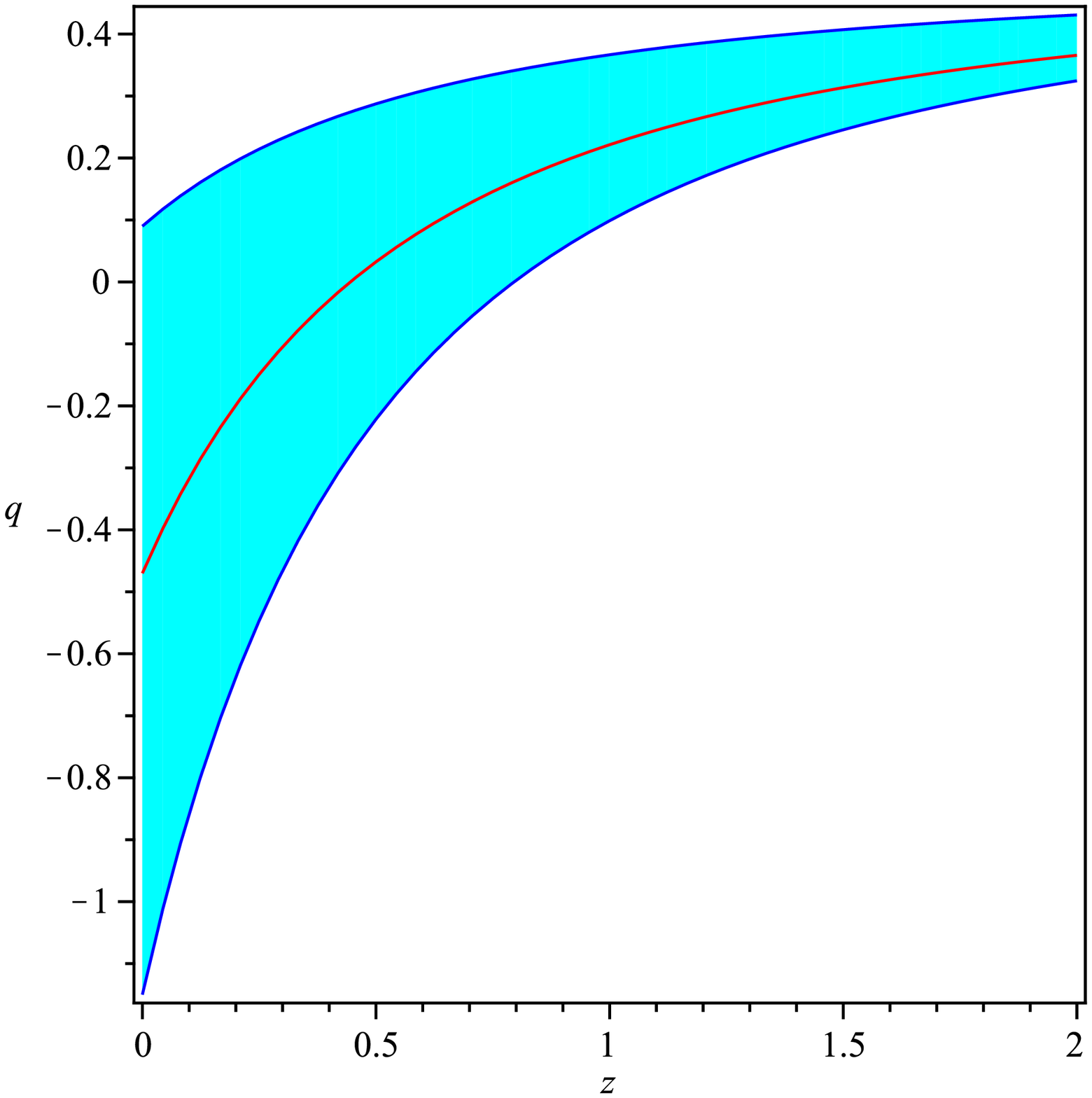}
                \caption{Hubble}
                \label{fig:Hubble}
        \end{subfigure}%
        ~ 
        \begin{subfigure}[b]{0.3\textwidth}
                \centering
                \includegraphics[width=\textwidth]{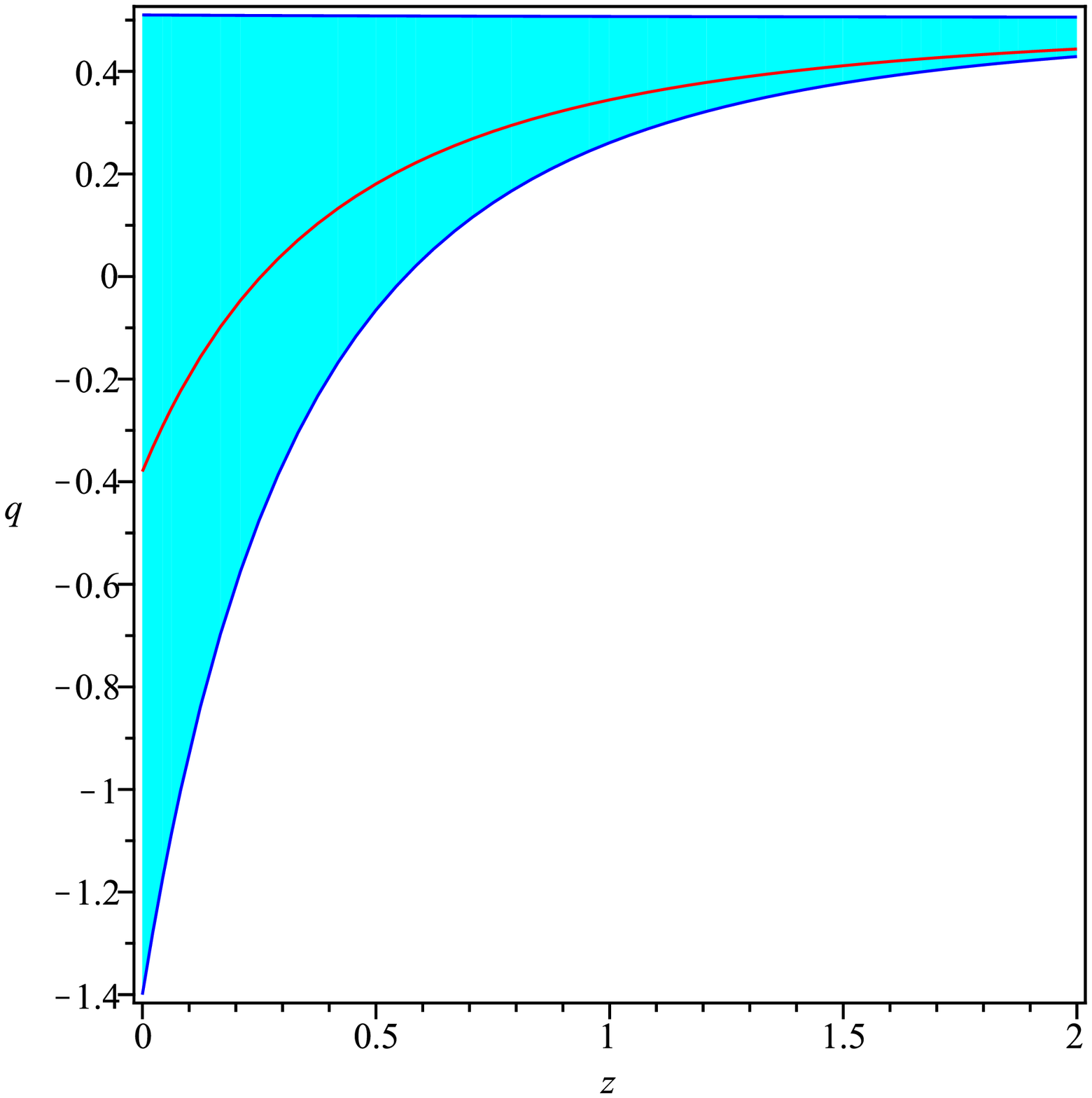}
                \caption{CMB}
                \label{fig:CMB}
        \end{subfigure}
        ~ 
        \begin{subfigure}[b]{0.3\textwidth}
                \centering
                \includegraphics[width=\textwidth]{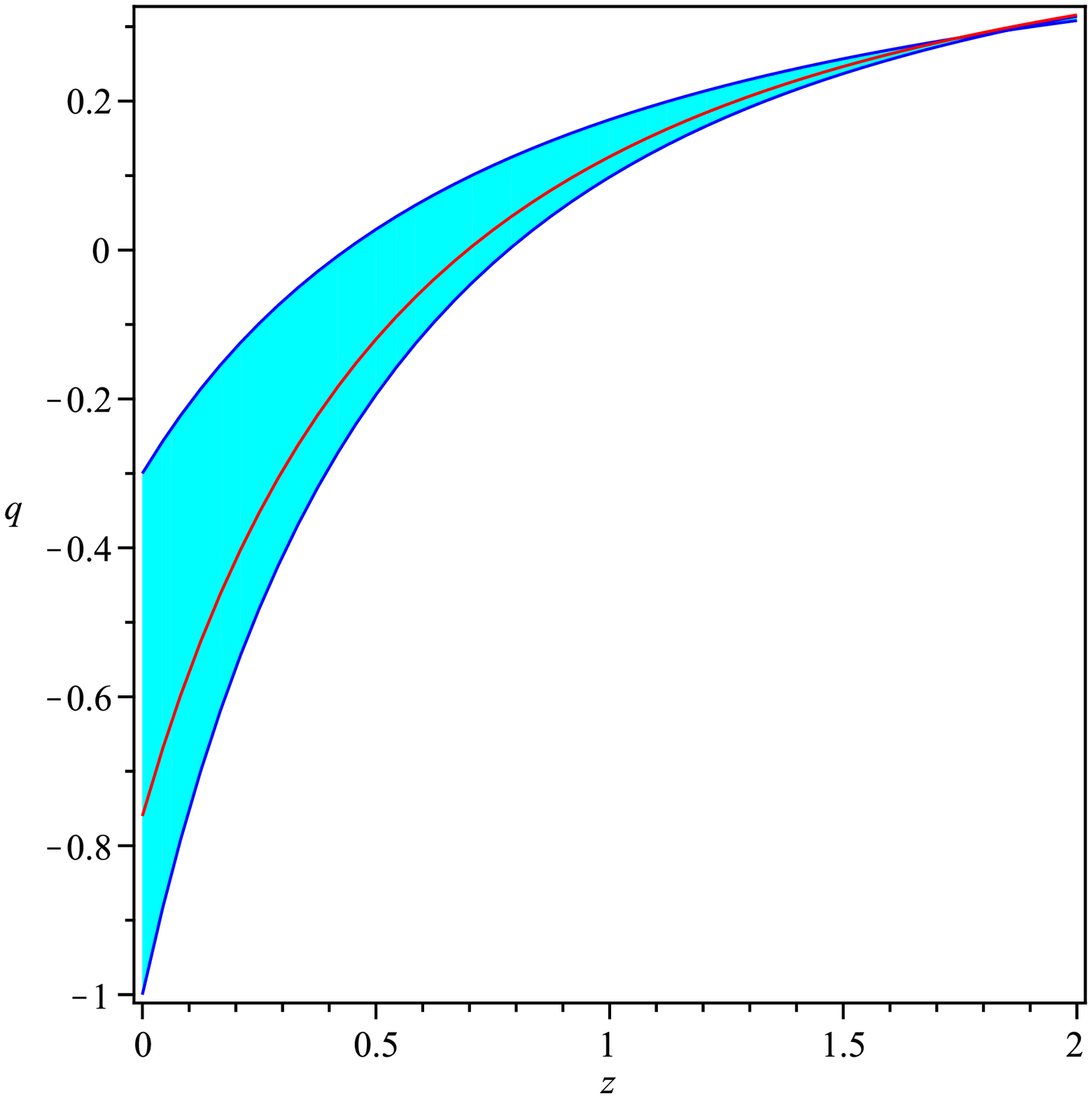}
                \caption{H+CMB}
                \label{fig:H+CMB}
        \end{subfigure}
        \caption{Evolutions of the decelerated parameter $q$ versus red shift z, which are constrained according to: (a) Hubble, (b) CMB, and (c) H + CMB combination data}\label{fig:Observational data}
\end{figure}
\begin{figure}
        \centering
        \begin{subfigure}[b]{0.3\textwidth}
                \centering
                \includegraphics[width=\textwidth]{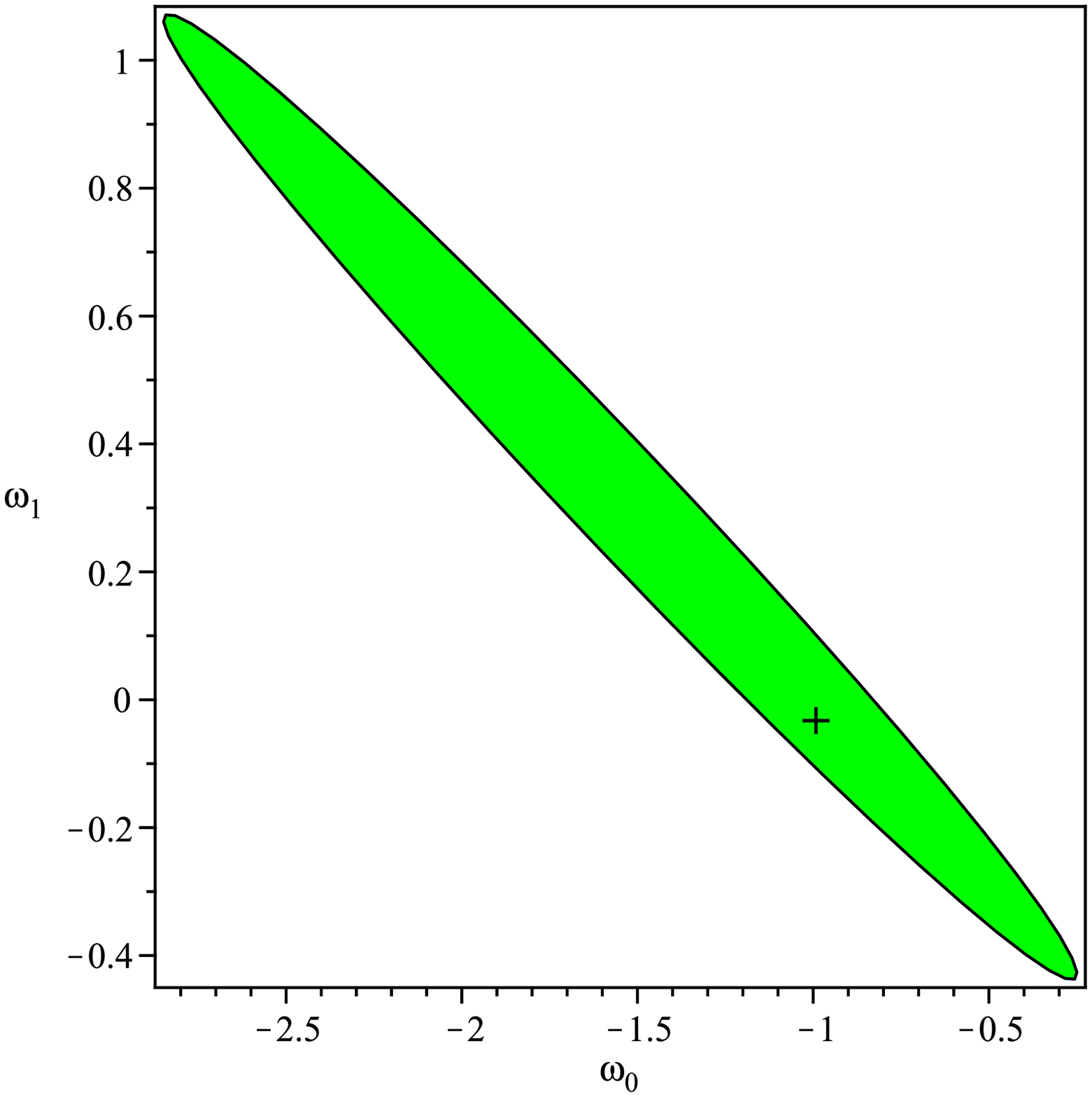}
                \caption{Hubble}
                \label{fig:Hubble}
        \end{subfigure}%
        ~ 
        \begin{subfigure}[b]{0.3\textwidth}
                \centering
                \includegraphics[width=\textwidth]{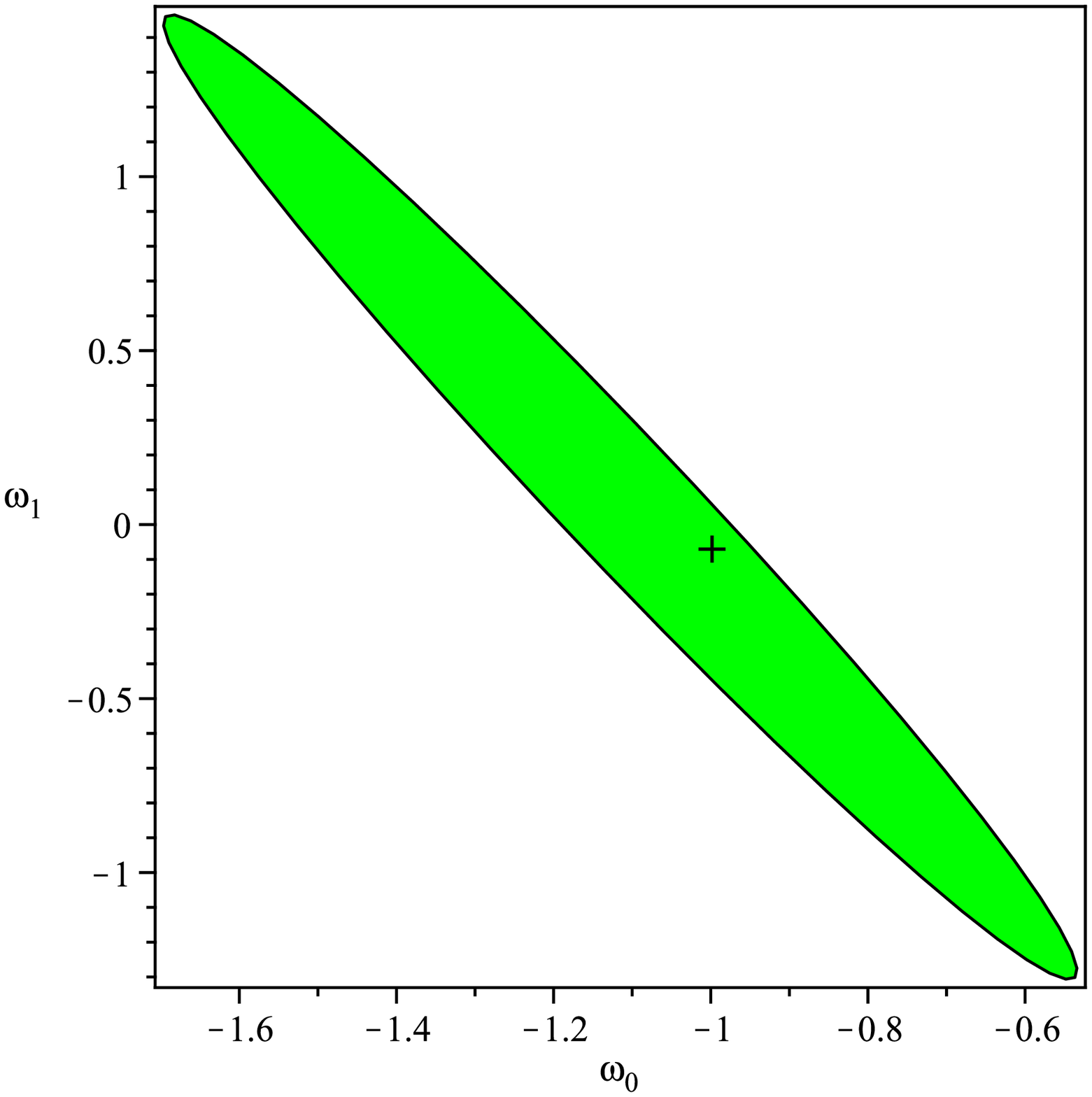}
                \caption{CMB}
                \label{fig:CMB}
        \end{subfigure}
        ~ 
        \begin{subfigure}[b]{0.3\textwidth}
                \centering
                \includegraphics[width=\textwidth]{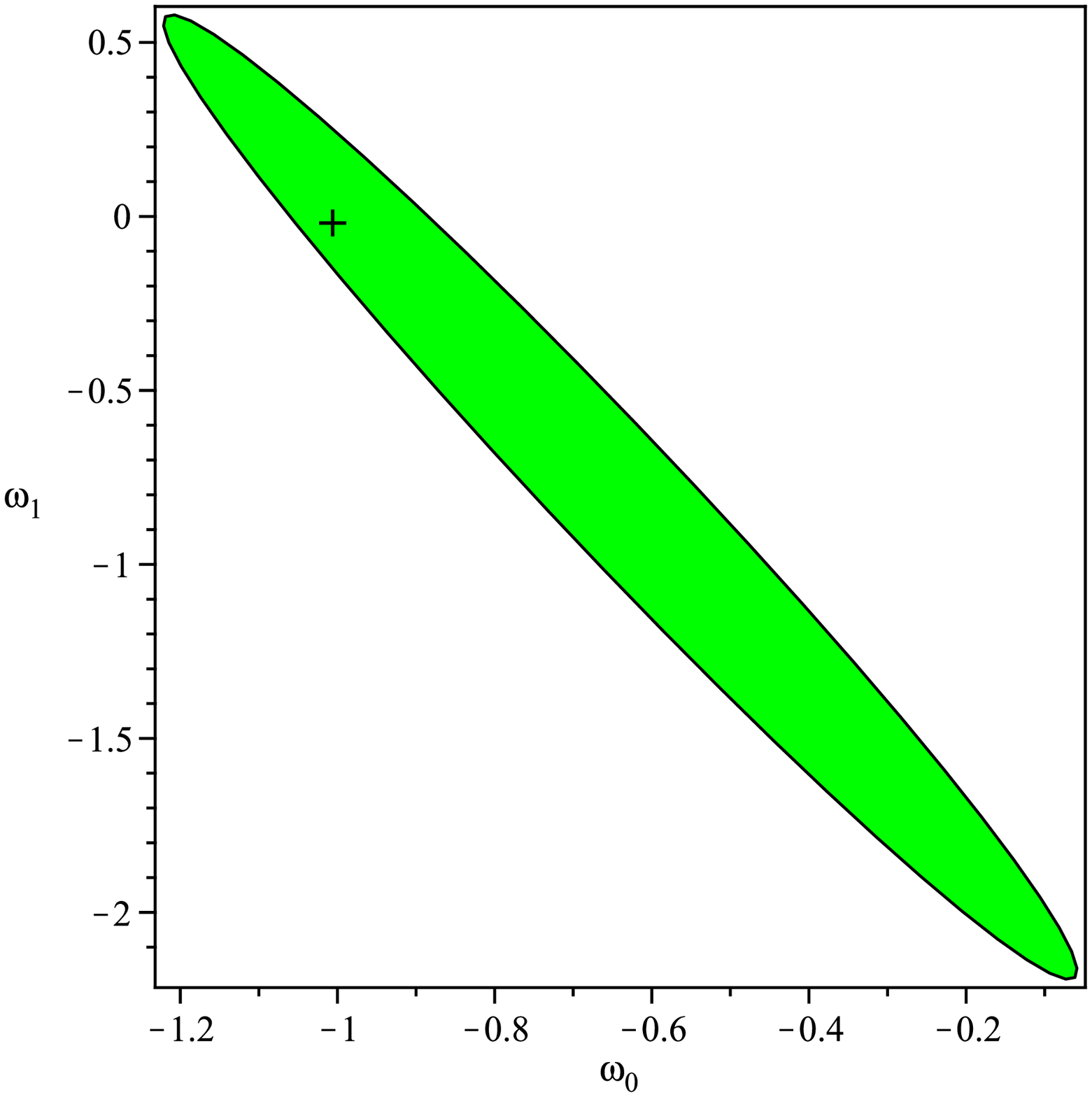}
                \caption{H+CMB}
                \label{fig:H+CMB}
        \end{subfigure}
        \caption{Constraints on the dark energy equation of state $\omega^{X}=\omega_{0}+\omega_{1}\frac{z}{1+z}$ obtained by: (a) Hubble, (b) CMB, and (c) H + CMB combination data. A cross sing in figures shows cosmological constant corresponds to $\omega^{X}=-1$.}\label{fig:Observational data}
\end{figure}
The best fit transition redshift $z_{\ast}$ for the case of combined constraint at the $1\sigma$ error is
$0.60^{+0.20}_{-0.10}$. It is worth to mention that the constraints to the deceleration parameter $q$ and the dark energy equation of state are looser in 1$\sigma$ error by using Hubble data than CMB data, which may be attributed to a fewer data points from Hubble data.\\

The variations of the DP parameter $q$ and the EoS parameter $\omega^{X}$ are plotted in fig. 4. Here we have used the best fit values of
$a$, $b$, $\omega_{0}$, and $\omega_{1}$ from Table 2.
\begin{figure}[htbp]
\centering
\includegraphics[width=10cm,height=10cm,angle=0]{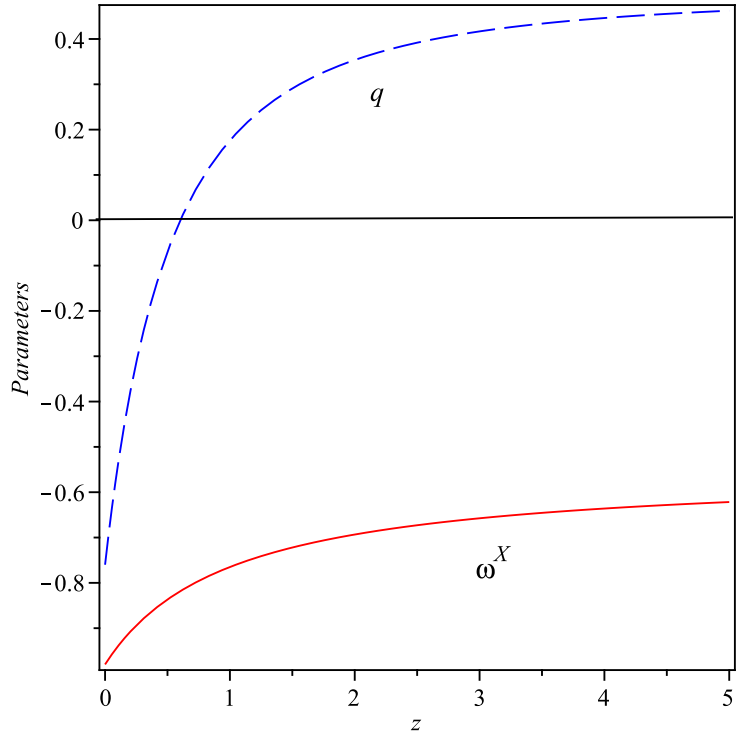}
\caption{The plot of DP parameter $q$ and $\Omega^{X}$ versus
redshift ($z$) for $a=1.26,~
b=1.96,~\omega_{0}=-0.98$, and $\omega_{0}=0.43$. These values are taken from the combined data in Table 2.}
\end{figure}
\section*{Acknowledgments}
This work has been supported by a research fund from the Mahshahr Branch of Islamic Azad University under the
project entitled ``The role of scalar fields in the study of dark energy". Author would like to thank S. Darvishi for his valuable help in calculating some parameters of table 2.



\begin{thebibliography}{000}
\bibitem {ref1}
S Perlmutter, {\it et al, Astrophys. J.} {\bf 517} 565 (1999)
\bibitem {ref2}
A G Riess {\it et al, Astron. J.} {\bf 116}, 1009 (1998)
\bibitem {ref3}
A G Riess {\it et al, Astrophys. J.} {\bf 560}, 49 (2001)
\bibitem {ref4}
J L Tonry {\it et al, Astrophys. J.} {\bf 594}, 1 (2003)
\bibitem {ref5}
M Tegmark {\it et al, Phys. Rev.} {\bf D69}, 103501 (2004)
\bibitem {ref6}
C L Bennet {\it et al, Astrophys. J. Suppl.} {\bf 148}, 1 (2003)
\bibitem {ref7}
D N Spergel {\it et al, Astrophys. J. Suppl.} {\bf 148}, 175 (2003)
\bibitem {ref8}
K Abazajian {\it et al, Astron. J.} {\bf 128}, 502 (2004)
\bibitem {ref9}
B Ratra and P. J. E. Peebles, {\it Phys. Rev.} {\bf D37}, 3406 (1988)
\bibitem {ref10}
C Wetterich, {\it Nucl. Phys.} {\bf B302}, 668 (1988)
\bibitem {ref11}
M S Turner and M. J. White, {\it Phys. Rev.} {\bf D56}, 4439 (1997)
\bibitem {ref12}
R R Caldwell, R. Dave and P. J. Steinhardt, { \it Phys. Rev. Lett.} {\bf 80}, 1582 (1998)
\bibitem {ref13}
A R Liddle and R. J. Scherrer, {\it Phys. Rev.} {\bf D59}, 023509 (1998)
\bibitem {ref14}
P J Steinhardt, L. M. Wang and I. Zlatev, {\it Phys. Rev.} {\bf D59}, 123504 (1999)
\bibitem {ref15}
R R Caldwell, {\it Phys. Lett.} {\bf B545} , 23 (2002)
\bibitem {ref16}
B Feng, X L Wang, and X Zhang, {\it Phys. Lett.} {\bf B607}, 35 (2005)
\bibitem {ref17}
A Picon, T Damour and V Mukhanov, {\it Phys. Lett.} {\bf B558}, 209 (1999)
\bibitem {ref18}
M Malquarti, E J Copeland, A R Liddle and M Trodden, {\it Phys. Rev.} {\bf D67}, 123503 (2003)
\bibitem {ref19}
S K Srivastava, {\it Phys. Lett.} {\bf B619}, 1 (2005)
\bibitem {ref20}
U Alam, V Sahni, T D Saini and A A Starobinsky, {\it Mon. Not. Roy. Astron. Soc.} {\bf 344}, 1057 (2003)
\bibitem {ref21}
A R Cooray and D. Huterer, {\it Astrophys. J.} {\bf 513}, L95 (1999)
\bibitem {ref22}
M Chevallier, D Polarski, {\it Int. J. Mod. Phys.} {\bf D10}, 213 (2001)
\bibitem {ref23}
E V Linder, {\it Phys. Rev. Lett.} {\bf 90}, 091301 (2003)
\bibitem {ref24}
A G Riess, {\it et al, Astrophys. J.} {\bf 659}, 98 (2007)
\bibitem {ref25}
L Xu, H Liu and Y Ping, {\it Int. Jour. Theor. Phys.} {\bf 45}, 869(2006)
\bibitem {ref26}
N Banerjee, S Das, {\it Gen.Rel.Grav.} {\bf 37}, 1695 (2005)
\bibitem {ref27}
M Goliath and G F R Ellis, {\it Phys. Rev} {\bf D60}, 032502 (1999)
\bibitem {ref28}
B Saha, {\it Mod. Phys. Lett.} {\bf A20}, 2127 (2005)
\bibitem {ref29}
E Komatsu, {\it et al, Astrophys. J. Suppl.} {\bf 192}, 18 (2011)
\bibitem {ref30}
L Xu, H Liu and Y Ping, {\it Int. J. Theor. Phys.} {\bf45}, 843 (2006)
\bibitem {ref31}
R Jimenez, L Verde, T Treu and D Stern, {\it Astrophys. J.} {\bf 593}, 622 (2003)
\bibitem {ref32}
R G Abraham, {\it et al, Astron. J.} {\bf 127}, 2455 (2004)
\bibitem {ref33}
T Treu, {\it et al, Mon. Not. Roy. Astron. Soc.} {\bf 308},1037 (1999)
\bibitem {ref34}
T Treu, {\it et al, Mon. Not. Roy. Astron. Soc.} {\bf 326}, 221 (2001)
\bibitem {ref35}
Simon J, {\it et al, Phys. Rev.} {\bf D71}, 123001 (2005)
\bibitem {ref36}
O Luongo, {\it Mod. Phys. Lett.} {\bf A26}, 1459 (2011)
\end{thebibliography}
\end{document}